\documentclass{article}
\usepackage{graphicx}
\begin{document}

\rightline{UMISS-HEP-2010-02}
\bigskip
\centerline{\bf \huge Magnetic Field Expansion Out of a Plane:}
\smallskip
\centerline{\bf \huge Application to Inverse Cyclotron Muon Cooling}
\bigskip



\centerline{\bf \large T. L. Hart\footnote{tlh@fnal.gov} and   D. J. Summers, U.~Mississippi-Oxford, University, MS 38677, USA}
\bigskip

\centerline{\bf \large K. Paul, Tech-X Corporation, Boulder, CO 80303, USA}
\medskip



\begin{abstract}
In studies of the dynamics of charged particles in a cyclotron magnetic field,
the specified field is generally $B_{z}$ in the $z = 0$ midplane where $B_{r}$
and $B_{\theta}$ are zero.  $B_r(r,\theta,z)$ and $B_{\theta}(r,\theta,z)$ are simple to
determine through a linear expansion which assumes that $B_z$ is 
independent of $z$.  But, an expansion to only first order may not be sufficient for
orbit simulations at large $z$.  This paper reviews the
expansion of a specified $B_{z}(r,\theta,z=0)$ out of the $z=0$ midplane to 
arbitrary order, and shows simple examples worked out to 4th order.
\end{abstract}

\leftline{{\bf Keywords}: muon ionization cooling, neutrino factory, muon collider} 


\section{Introduction}
Many programs and software packages 
simulate
orbits of charged particles in magnetic fields as noted in Table 1.   
Some use a single reference orbit and represent the magnetic field as a series of matrices, one for each magnetic element.
Other programs allow more realistic spiral orbits by using magnetic field equations or interpolated magnetic field maps for point to point 
Runge-Kutta track integrators.  Some programs such as COSY Infinity use exact analytical determination of magnetic field derivatives; others use 
numerical differentiation.  
An 
important preliminary step before simulation of fields produced by magnets
is investigation of orbit dynamics in an idealized magnetic field which
satisfies Maxwell's equations in vacuum to sufficient order.  Magnetic fields
for cyclotrons \cite{Livingood} are most conveniently expressed in cylindrical 
($r$,$\theta$,$z$) 
coordinates in which the $z$ direction is the cyclotron rotation axis.
In many applications, an ideal field is specfied in the cyclotron median plane
which is the $z=0$ plane where $B_r$ and $B_{\theta}$ vanish.  Determination
of ($B_r$,$B_\theta$,$B_z$) for $z \neq 0$ to sufficient order (beyond 1st
order) is necessary for detailed orbit and dynamics studies \cite{Jeon}.  Reduction of
the phase space of a muon beam through ionization cooling in an inverse cyclotron 
is being researched 
\cite{anti} as part of the muon R\&D for possible future
facilities such as a neutrino factory and muon collider 
\cite{nfmc}.
\medskip

\begin{table}[h!]
\begin{center}
\renewcommand{\arraystretch}{1.10}
\tabcolsep=1.2mm
\begin{tabular}{lcccc} \hline
                        &  Tracks                                                & Tracking                 &  Tracking                  &                     \\
Program        &  in Matter     &  Volume &                                 Method    & Reference \\ \hline
G4Beamline (GEANT4) &  yes                          &  unlimited               &  Point to Point           & \cite{Roberts} \\  
VORPAL                                                       & yes  &   unlimited               &     Point to Point                            & \cite{Cary} \\
ICOOL             &  yes                          &  near a line             & Point to Point           & \cite{Fernow} \\
Methodical Accelerator Design (MAD-X) & no & near a line & Matrices or Points   & \cite{MAD} \\
SYNCH                                                        & no &  near a line                   & Matrices or Points                                & \cite{Garren} \\
Zgoubi                                                          & no &   unlimited                  &         Point to Point                        & \cite{Meot} \\
COSY Infinity                                               & yes  &   unlimited                 &     Transfer Maps or Points                               & \cite{Berz} \\
CYCLOPS  and GOBLIN                                                     & no    &    near a plane                &  Point to Point                                 & \cite{Gordon}  \\
OptiM                                                           &  yes   &      near a line            &            Matrix Elements                     & \cite{Lebedev} \\ \hline

\end{tabular}
\caption{Computer programs for tracking charged particles in magnetic fields and designing accelerators.}   
\end{center}
\end{table}

\section{Expansion of Field Out of the Midplane}
The importance of correct field expansion for reliable particle
tracking simulation and beam dynamics analysis has been recognized,
and the expansion around the reference orbit of a bent-solenoid
geometry has been calculated \cite{bent-sol}. Expansions to eighth order are done with the aid of Mathematica.
A static magnetic field can be expanded out of a plane because it must  satisfy the source-free Maxwell equations in
vacuum, $\nabla \cdot \bf{B} \rm = 0$ and $\nabla \times \bf{B} \rm = 0$.  
Only ${\partial B_x}/{\partial y} \! =  \! {\partial B_y}/{\partial x}$ constrains fields in the plane, so any function can be used for $B_z$.
One gets the three field components out of the plane as follows

\[ \frac{\partial B_z}{\partial z} = - \frac{\partial B_x}{\partial x}  \, - \, \frac{\partial B_y}{\partial y} \qquad \qquad \frac{\partial B_x}{\partial z} 
=  \frac{\partial B_z}{\partial x}
\qquad \qquad \frac{\partial B_y}{\partial z} =  \frac{\partial B_z}{\partial y}\]

The expansion of a magnetic field specified in a plane has been addressed by Zgoubi
for cartesian coordinates \cite{zguser}.  Zgoubi uses Taylor expansion with  $\nabla \cdot \bf{B} \rm = 0$ and $\nabla \times \bf{B} \rm = 0$.
Here we work out a solution in cylindrical coordinates to arbitrary order.
Median plane antisymmetry is assumed so that

\[B_{r}(r,\theta,0) = 0 \qquad B_{\theta}(r,\theta,0) = 0\]

\[B_{r}(r,\theta,z) = -B_{r}(r,\theta,-z) \qquad B_{\theta}(r,\theta,z) = -B_{\theta}(r,\theta,-z) \qquad B_{z}(r,\theta,z) = B_{z}(r,\theta,-z)\]

With
$B \equiv B_{z}(r,\theta,z=0)$ the magnetic field is
\[B_{r} = z \frac{\partial B}{\partial r} - 
\frac{z^{3}}{6} \frac{\partial}{\partial r} \left(\frac{\partial^{2}B}{\partial r^{2}} +
\frac{1}{r}\frac{\partial B}{\partial r} + 
\frac{1}{r^2}\frac{\partial^{2} B}{\partial \theta^{2}}\right) + \ldots\]

\[B_{\theta} = \frac{z}{r} \frac{\partial B}{\partial \theta} - 
\frac{z^{3}}{6} \frac{1}{r} \frac{\partial}{\partial \theta}\left(\frac{\partial^{2}B}{\partial r^{2}} +
\frac{1}{r}\frac{\partial B}{\partial r} + 
\frac{1}{r^2}\frac{\partial^{2} B}{\partial \theta^{2}}\right) + \ldots\]

\[B_{z} = B - 
\frac{z^{2}}{2} \left(\frac{\partial^{2}B}{\partial r^{2}} +
\frac{1}{r}\frac{\partial B}{\partial r} + 
\frac{1}{r^2}\frac{\partial^{2} B}{\partial \theta^{2}}\right) + \ldots\]

which is

\[B_{r} = z \frac{\partial B}{\partial r} + 
\frac{\partial}{\partial r} \left[\sum_{n=1}^{\infty}(-1)^{n}\frac{z^{2n+1}}{(2n+1)!}\left(\frac{\partial^{2}}{\partial r^{2}} +
\frac{1}{r}\frac{\partial}{\partial r} + 
\frac{1}{r^2}\frac{\partial^{2}}{\partial \theta^{2}}\right)^{n} B\right]\]

\[B_{\theta} = \frac{z}{r} \frac{\partial B}{\partial \theta} + 
\frac{1}{r}\frac{\partial}{\partial \theta} \left[\sum_{n=1}^{\infty}(-1)^{n}\frac{z^{2n+1}}{(2n+1)!}\left(\frac{\partial^{2}}{\partial r^{2}} +
\frac{1}{r}\frac{\partial}{\partial r} + 
\frac{1}{r^2}\frac{\partial^{2}}{\partial \theta^{2}}\right)^{n} B\right]\]

\[B_{z} = B + 
\left[\sum_{n=1}^{\infty}(-1)^{n}\frac{z^{2n}}{(2n)!}\left(\frac{\partial^{2}}{\partial r^{2}} +
\frac{1}{r}\frac{\partial}{\partial r} + 
\frac{1}{r^2}\frac{\partial^{2}}{\partial \theta^{2}}\right)^{n} B\right]\]
A rigorous proof of these expansions is provided in the final section of this
paper.

In most implementations of simulating a magnetic field, the expansions of
$(B_{r}, B_{\theta}, B_{z})$ cannot be expressed in closed form.  In such
cases, only a finite number of terms of the magnetic field expansion can be 
used.  The order of the expansion is defined as the highest power of $z$ in
the summation.  To odd (even) order $m$, 
$B_{r}$ and $B_{\theta}$ have $\frac{m+1}{2}$ (
$\frac{m}{2}$) terms while $B_z$ has $\frac{m+1}{2}$ ($\frac{m+2}{2}$) terms.
When $\bf{B}$ is expanded to odd order, $\nabla \times \bf{B} \rm = 0$ and 
$\nabla \cdot \bf{B} \rm \neq 0$; when $\bf{B}$ is expanded to even order, 
$\nabla \cdot \bf{B} \rm = 0$ and $\nabla \times \bf{B} \rm \neq 0$.
An even order expansion with     
$\nabla \cdot \bf{B} \rm = 0$ 
is needed to satisfy the Hamiltonian and allow a symplectic map \cite{Forest}.

\section{Example Fields to 4th Order}

This section shows some example midplane fields $B_z(r,\theta,z=0)$ expanded
to 4th order to obtain $B_r(r,\theta,z)$, $B_\theta(r,\theta,z)$, and 
$B_z(r,\theta,z)$.  These example fields have fairly simple radial and 
azimuthal dependences and were checked with Mathemtica.  The 
4th order expansion of a more complex field,
one with a constant radial field index, $k$ and spiral angle, $\alpha$, such
as $B_z(z=0) = cr^k(1 - f\sin[N(\theta - \tan\alpha\ln(r/r_0))])$, yields
pages of output so that the use of a program such as Mathematica is
necessary.  Mathematica includes a command which produces output in 
{\tt FORTRAN} or 
{\tt C} syntax which can be copied directly into a routine which 
generates a grid of magnetic field points.  Example Mathematica and
{\tt FORTRAN} files are presented in Appendices A and B. 

\subsection{$B$ Only Dependent on Radius (No Sectors)}

Is this case, $B(r,\theta,z=0)$ is $B(r,z=0)$ so that 
\newline
\[B_{r} = z \frac{\partial B}{\partial r} - 
\frac{z^{3}}{6} \frac{\partial}{\partial r} \left(\frac{\partial^{2}B}{\partial r^{2}} + \frac{1}{r}\frac{\partial B}{\partial r}\right) \]
\newline
\[B_{\theta} = 0 \]
\newline
\[B_{z} = B(r,z=0) - 
\frac{z^{2}}{2} \left(\frac{\partial^{2}B}{\partial r^{2}} +
\frac{1}{r}\frac{\partial B}{\partial r}\right) +
\frac{z^4}{24}\left(\frac{\partial^4 B}{\partial r^4} + 
   \frac{\partial^2}{\partial r^2}\left(\frac{1}{r} \frac{\partial B}{\partial r}\right) + 
   \frac{1}{r} \frac{\partial^3 B}{\partial r^3} +
   \frac{1}{r} \frac{\partial}{\partial r} \left( \frac{1}{r} \frac{\partial B}{\partial r} \right) \right)
\]

An example of such a field is a non-sectored field in the middle of Helmoltz 
coils.  At $z=0$, the $B_z$ is maximal at $r=0$ and decreases very slowly with
radius so that $\partial B / \partial r$ and higher derivatives are small,
but still large enough to provide weak focusing in $r$ and $z$.

\subsection{Azimuthal Field Sectors:  No Spiral Angle nor Midplane Radial Dependence}

For this field $B(r,\theta,z=0)$ is $B(\theta,z=0)$:  an example is
$B=B_0(1-f\sin(N\theta))$ so that
\newline
\[B_{r} = \frac{1}{3}\left(\frac{z}{r}\right)^{3}\left(fB_{0}N^2 \sin(N\theta)\right) \]
\newline
\[B_{\theta} = -\frac{z}{r}(f B_0 N \cos(N\theta)) - 
\frac{1}{6} \left(\frac{z}{r} \right)^3 (f B_0 N^3 \cos(N\theta))    \]
\newline
\[B_{z} = B_0 (1 - f \sin(N \theta)) - 
\frac{1}{2} \left(\frac{z}{r} \right)^2 (f B_0 N^2 \sin(N\theta)) +
\frac{1}{24} \left(\frac{z}{r} \right)^4 (f B_0 N^2 \sin(N\theta))(4 - N^2) \]

A sectored field such as this provides stronger focusing in $z$ than a
non-sectored field.

\subsection{Azimuthal Field Sectors:  Constant Spiral Angle, No Radial Field Index}

For this field, $B(r,\theta,z=0)$ is $B(\theta,z=0)$:  a field with
constant spiral angle $\alpha$ is \newline 

$B=B_0(1-f\sin[N(\theta-\tan\alpha\ln(r/r_0))])$ so that
\newline
\[B_{r} = \frac{z}{r}(f B_0 N \tan\alpha) (\cos[N(\theta-\tan\alpha\ln(r/r_0))])+ \]
\[
\frac{1}{6}\left(\frac{z}{r}\right)^{3}(fB_{0}N^2\sec^2\alpha)(N\tan\alpha\cos[N(\theta-\tan\alpha\log(r/r_0))] + 2\sin[N(\theta-\tan\alpha\log(r/r_0))]) \]
\newline
\[B_{\theta} = -\frac{z}{r}(f B_0 N) (\cos[N(\theta-\tan\alpha\ln(r/r_0))]) - 
\frac{1}{6} \left(\frac{z}{r} \right)^3 (f B_0 N^3 \sec^2\alpha) (\cos[N(\theta-\tan\alpha\ln(r/r_0))])    \]
\newline
\[B_{z} = B_0 (1 - f \sin[N(\theta-\tan\alpha\ln(r/r_0))]) - \]
\[
\frac{1}{2} \left(\frac{z}{r} \right)^2 (f B_0 N^2 \sec^2\alpha)(\sin[N(\theta-\tan\alpha\ln(r/r_0))]) + \]
\[
\frac{1}{24} \left(\frac{z}{r} \right)^4 (f B_0 N^2 \sec^2\alpha)\big\{-N^2 \sec^2\alpha\sin[N(\theta-\tan\alpha\ln(r/r_0))] + \]
\[
4(\sin[N(\theta-\tan\alpha\ln(r/r_0))] + N \tan\alpha \cos[N(\theta-\tan\alpha\ln(r/r_0))]\big\}
\]

The spiraled magnetic field provides stronger focusing in $z$ than the
non-spiraled sectored field.  When $\alpha$ = 0, the magnetic field components 
reduce to those of
the previous azimuthally sectored field with no spiral angle nor midplane 
radial dependence.

\subsection{Azimuthal Field Sectors:  No Spiral Angle, Constant Radial Field Index}

\[k \equiv \frac{r}{B_{z}(z=0)} \frac{\partial B_{z}(z=0)}{\partial r}\]

For this field, $B(r,\theta,z=0) = cr^{\,k}(1 - f \sin(N \theta))$.  Then
\newline
\[B_r = \left(\frac{z}{r}\right)(cr^{\,k})k(1-f\sin(N\theta)) - 
\frac{1}{6}\left(\frac{z}{r}\right)^3(cr^{\,k})(k-2)(fN^2\sin(N\theta)+k^2(1-f\sin(N\theta)))\]
\newline
\[B_\theta = - \left(\frac{z}{r}\right)(cr^{\,k})fN\cos(N\theta) - 
\frac{1}{6}\left(\frac{z}{r}\right)^3(cr^{\,k})(fN\cos(N\theta))(N^2-k^2)\]
\newline
\[B_{z} = c r^{\,k} (1-f \sin (N \theta)) - \]
\[   \frac{1}{2} \left(\frac{z}{r}\right)^2 (c r^{\,k}) (f N^2 \sin(N \theta) + 
   k^2(1 - f\sin(N \theta))) + \]
\[   \frac{1}{24} \left(\frac{z}{r}\right)^4 (cr^{\,k})\left[
   fN^2\sin(N\theta)(2(k^2-2k+2)-N^2) + (k-2)^2k^2(1-f\sin(N\theta))
   \right]
\]

\section{Proof of Expansion Out of the Plane}

Let 
\newline
${\bf B}(r,\theta,z) = \hat{r}B_r(r,\theta,z) + \hat{\theta}B_\theta(r,\theta,z)+ \hat{z}B_z(r,\theta,z) \equiv 
{\bf B}_\bot(r,\theta,z) + \hat{z}B_z(r,\theta,z),$
\newline
\newline
${\bf B}_\bot(r,\theta,z) =$ \(\sum_{k=0}^{\infty}{\bf A}_k(r,\theta)z^k \),
\newline
$B_z(r,\theta,z) =$ \(\sum_{k=0}^{\infty}B_k(r,\theta)z^k \),
\newline
\newline
For any vector field {\bf V} and scalar field $S$,
\newline
$\vec{\nabla}_\bot \cdot {\bf V}_\bot = \frac{1}{r} \frac{\partial}{\partial r}(r V_{r}) + \frac{1}{r}\frac{\partial V_\theta}{\partial \theta}$,
\newline
$\vec{\nabla}_\bot {S} = \hat{r}\frac{\partial S}{\partial r} + \hat{\theta}\frac{1}{r}\frac{\partial S}{\partial \theta}$.
\newline
$\vec{\nabla}_\bot^2 S \equiv \vec{\nabla}_\bot \cdot (\vec{\nabla}_\bot S) =
\frac{\partial^2 S}{\partial r^2} + \frac{1}{r} \frac{\partial S}{\partial r} +
\frac{1}{r^2}\frac{\partial^2 S}{\partial \theta^2}$
\newline
\newline
It is necessary and sufficient to show that the specification of \newline
$B_{0}(r,\theta), \vec{\nabla} \cdot {\bf B} = 0, \vec{\nabla} \times {\bf B} = 0
\Rightarrow \newline
{\bf A}_{2n+1}(r,\theta) = \frac{(-1)^{n}}{(2n+1)!}\vec{\nabla}_\bot \left( (\vec{\nabla}_\bot^{2} )^{n} B_0 \right)$ for $n \geq 1$ and \newline 
$B_{2n}(r,\theta) = \frac{(-1)^{n}}{(2n)!} (\vec{\nabla}_\bot^{2})^n B_0$ for $n \geq 1$. 
\newline
\newline
Note that $B$ was defined as $B_0(r,\theta)$.
\newline
\newline
$\vec{\nabla}\cdot{\bf B} = 0 \Rightarrow 
\newline
\left(\vec{\nabla}_\bot + \hat{z}\frac{\partial}{\partial z}\right)\cdot
\left({\bf B}_\bot + \hat{z}B_z\right) = \newline
\vec{\nabla}_\bot\cdot{\bf B}_\bot + \frac{\partial B_z}{\partial z} = \newline
\sum_{k=0}^{\infty}(\vec{\nabla}_\bot \cdot{\bf A}_k) z^k + \sum_{k=1}^{\infty}k B_k z^{k-1} = \newline
\sum_{k=0}^{\infty}\left[(\vec{\nabla}_\bot \cdot {\bf A}_k) + (k + 1)B_{k+1}\right]z^k = 0 \Rightarrow \newline
(\vec{\nabla}_\bot \cdot {\bf A}_k) + (k + 1)B_{k+1} = 0 \Rightarrow \newline
B_{k+1} = -\frac{1}{k+1}(\vec{\nabla}_\bot \cdot {\bf A}_k)$ for $k = 0, 1, 2, \ldots$
\newline
\newline
$\vec{\nabla}\times{\bf B} = 0 \Rightarrow \newline
\left(\vec{\nabla}_\bot + \hat{z}\frac{\partial}{\partial z}\right)\times
\left({\bf B}_\bot + \hat{z}B_z\right) = \newline
(\vec{\nabla}_\bot \times {\bf B}_\bot) + \left(\vec{\nabla}_\bot \times (\hat{z} B_z)\right) +
\left(\left(\hat{z} \frac{\partial}{\partial z} \right) \times \vec{{\bf B}}_\bot \right) = \newline
(\vec{\nabla}_\bot \times {\bf B}_\bot) - \hat{z} \times (\vec{\nabla}_\bot B_z) +
\hat{z} \times \left( \frac{\partial {\bf \vec{B}}_\bot}{\partial z} \right) = \newline(\vec{\nabla}_\bot \times {\bf B}_\bot) - \hat{z} \times \left(\vec{\nabla}_\bot B_z -
\frac{\partial {\bf \vec{B}}_\bot}{\partial z} \right) = 0. \newline
(\vec{\nabla}_\bot \times {\bf B}_\bot) \mbox{ is parallel to } \hat{z} \mbox { and } \hat{z} \times \left(\vec{\nabla}_\bot B_z -
\frac{\partial {\bf \vec{B}}_\bot}{\partial z} \right) \mbox{ is perpendicular to } \hat{z}$.  
\newline
Therefore, $(\vec{\nabla}_\bot \times {\bf B}_\bot) = 0$ and 
$\left(\vec{\nabla}_\bot B_z -
\frac{\partial {\bf \vec{B}}_\bot}{\partial z} \right) = 0$.
\newline
Then $\left(\vec{\nabla}_\bot B_z -
\frac{\partial {\bf \vec{B}}_\bot}{\partial z} \right) = 0 \Rightarrow
\sum_{k=0}^{\infty} (\vec{\nabla}_\bot B_k)z^k - \sum_{k=1}^{\infty} k {\bf A}_k z^{k-1} = 0 \Rightarrow \newline
\sum_{k=0}^{\infty}\left((\vec{\nabla}_\bot B_k) - (k+1){\bf A}_{k+1}\right)z^k = 0 \Rightarrow \newline
{\bf A}_{k+1} = \frac{1}{k+1}(\vec{\nabla}_\bot B_k)$ for $k = 0, 1, 2, \ldots$
\newline
\newline
\newline
$B_{k+1} = -\frac{1}{k+1}(\vec{\nabla}_\bot \cdot {\bf A}_k) \mbox{ and }
{\bf A}_{k+1} = \frac{1}{k+1}(\vec{\nabla}_\bot B_k) \Rightarrow \newline
B_{k+2} = -\frac{1}{(k+2)(k+1)}\left(\vec{\nabla}_\bot \cdot (\vec{\nabla}_\bot B_k)\right) = -\frac{1}{(k+1)(k+2)}\vec{\nabla}_\bot^2 B_k$
\newline
\newline
$k=0 \Rightarrow \newline
{\bf A}_1 = \vec{\nabla}_\bot B_0 \newline
B_2 = -\frac{1}{2}\vec{\nabla}^2_\bot B_0 \newline \newline
{\bf A}_0 = 0 \newline
B_1 = 0$
\newline
\newline
$k=1 \Rightarrow \newline
{\bf A}_2 = \frac{1}{2}\vec{\nabla}_\bot B_1 = 0 \newline
B_3 = -\frac{1}{(3)(2)}\vec{\nabla}^2_\bot B_1 = 0$ 
\newline 
\newline
$k=2 \Rightarrow \newline
{\bf A}_3 = \frac{1}{3}\vec{\nabla}_\bot B_2 = -\frac{1}{(3)(2)}\vec{\nabla}_\bot \left(\vec{\nabla}^2_\bot B_0 \right) = -\frac{1}{3!}\vec{\nabla}_\bot \left(\vec{\nabla}^2_\bot B_0 \right)\newline
B_4 = -\frac{1}{(4)(3)}\vec{\nabla}^2_\bot B_2 = \frac{1}{(4)(3)(2)}\vec{\nabla}^2_\bot\left(\vec{\nabla}^2_\bot B_0 \right) = \frac{1}{4!}\vec{\nabla}^2_\bot\left(\vec{\nabla}^2_\bot B_0 \right)$ 
\newline
\newline
Continuing recursively,
\newline
$k=2n-1 \Rightarrow \newline
{\bf A}_{2n} = 0 \mbox{ for } n \geq 1 \newline
B_{2n-1} = 0 \mbox{ for } n \geq 1 $
\newline
\newline
$k=2n \Rightarrow \newline
{\bf A}_{2n+1} = \frac{(-1)^{n}}{(2n+1)!}\vec{\nabla}_\bot \left((\vec{\nabla}^{2})^{n}_\bot B_0 \right) \mbox{ for } n \geq 1 \newline
B_{2n} = \frac{(-1)^{n}}{(2n)!}(\vec{\nabla}^{2})^n_\bot B_0
\mbox{ for } n \geq 1$
\newline
\newline
The proof is complete.

\section{Summary}

The procedure for expanding a specified $B_z(r,\theta,z=0)$ out of the 
$z=0$ midplane is described.  Examples of expansions out to 4th order are
included.  Fig.~1 shows an example of a magnetic field.
\medskip

\begin{figure}[h!]
    \includegraphics[width=100mm]{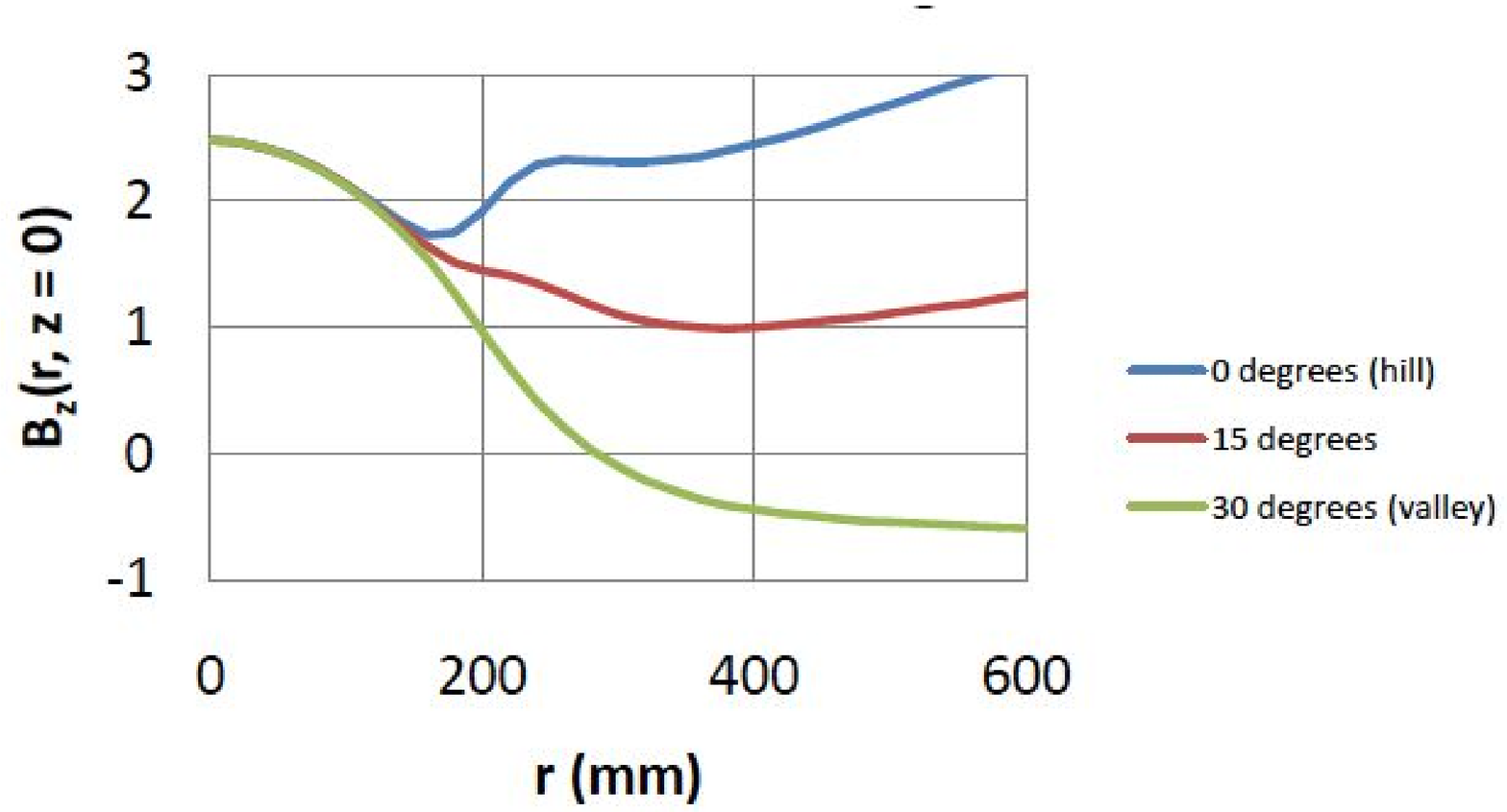}
    \includegraphics[width=66mm]{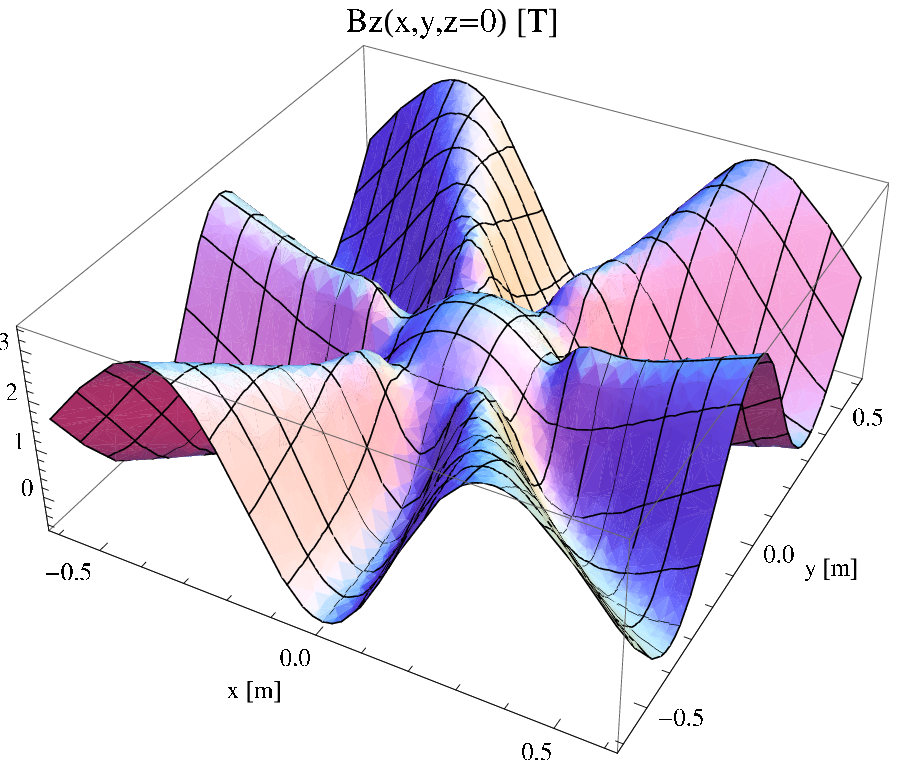}
\vspace{-3mm}    
    \caption{$B_z(x,y,z=0) = 1.77 r^{\,0.6} (1 + \sqrt{2} \cos{(6\theta)}) (1 + \tanh{(25(r - 0.2)})/2$ including the outer 6-sector focus field which merges into the
             inner magnetic bottle.    
             $B_z(x,y,z=0)$ ranges from -\,0.6 to 3.2 Tesla.
The inner bottle field is
generated by two $r$ = 0.2 m circular coils, located 0.2 m above and
below the midplane. The $B_z$ field at $r$ = 0, $z$ = 0 is 2.4 Tesla.}
    \label{bz0}
\end{figure}



%


This work was supported by National Science Foundation Award 757938,
DOE  grant DE-FG05-91ER40622, and DOE SBIR DE-FG02-08ER85044.
Many thanks to T.~Roberts, R.~Fernow, J.~Gallardo, S.~Berg, A.~Garren, M.~Berz, A.~Bogacz, and R.~Palmer for their advice and help.


\appendix

\section{Appendix:  Mathematica Example}

An example of a Mathematica file which evaluates the coefficients of 
$B_z(r,\theta,z=0) = c r^k (1 + f\cos(N\theta))$ up to 4th order
is shown.  This $B_z(r, \theta,z=0)$ is simpler than the field of 
Fig.~\ref{bz0} so that the example file and output are reasonably
short and clear.  The example
file also formats the expressions of the coefficients to {\tt FORTRAN}.
The next appendix is an example {\tt FORTRAN} routine which uses the output
{\tt FORTRAN} expressions from this file to generate a grid of magnetic field
points used by a G4Beamline input file (Appendix C).  This Mathematica file 
shows the output of each Mathematica command indented.

\begin{verbatim}
(* Set b as Bz(z = 0) with (r, t) as radius and theta. *)
b[r_,t_] = (c r^k)(1+f Cos[n ( t)])
  
     c r^k (1+f Cos[n t])

(* Set trans as d^2b/dr^2 + (1/r)db/r + (1/r^2)d^2b/dt^2. *)
trans[r_,t_] = D[b[r,t],{r,2}] + (1/r)D[b[r,t],{r,1}]+ (1/r^2)D[b[r,t],{t,2}]

     -c f n^2 r^(-2+k) Cos[n t]+c k r^(-2+k) (1+f Cos[n t])+c (-1+k) k 
     r^(-2+k)(1+f Cos[n t])

(* FullSimplify does an algebraic simplification *)
FullSimplify[%]

     c r^(-2+k) (k^2+f (k-n) (k+n) Cos[n t])

(* trans2 is a double application of the derivatives of trans. *)
trans2[r_,t_] = D[trans[r,t],{r,2}] + (1/r)D[trans[r,t],{r,1}] + 
(1/r^2)D[trans[r,t],{t,2}]

     -c f (-3+k) (-2+k) n^2 r^(-4+k) Cos[n t]+c (-3+k) (-2+k) k r^(-4+k) 
     (1+f Cos[n t])+c (-3+k) (-2+k) (-1+k) k r^(-4+k) (1+f Cos[n t])+
     (-c f k n^2 r^(-2+k) Cos[n t]-c f (-1+k) k n^2 r^(-2+k) Cos[n t]+
     c f n^4 r^(-2+k) Cos[n t])/r^2+1/r (-c f (-2+k) n^2 r^(-3+k) Cos[n t]+
     c (-2+k) k r^(-3+k) (1+f Cos[n t])+c (-2+k) (-1+k) k r^(-3+k) 
     (1+f Cos[n t]))

FullSimplify[%]

     c r^(-4+k) ((-2+k)^2 k^2+f (-2+k-n) (k-n) (-2+k+n) (k+n) Cos[n t])

(* Set field components, apply simplifications, convert to *)
(* FORTRAN format. *)

(* Use o1, o2, o3, and o4 as switches to set order of expansion. *)
(* For 2nd order, set o1 = 1, o2 = 1, o3 = o4 = 0. *)

(* Set the 0th, 2nd, and 4th order terms of Bz(z). *)
bz0[r_,t_] = b[r,t]

     c r^k (1+f Cos[n t])

FullSimplify[%]

     c r^k (1+f Cos[n t])

FortranForm[%]

     c*r**k*(1 + f*Cos(n*t))

bz2[r_,t_] = -(o2)(z^2/2)trans[r,t]

     -(1/2) o2 z^2 (-c f n^2 r^(-2+k) Cos[n t]+c k r^(-2+k) (1+f Cos[n t])+
     c (-1+k) k r^(-2+k) (1+f Cos[n t]))

FullSimplify[%]

     -(1/2) c o2 r^(-2+k) z^2 (k^2+f (k-n) (k+n) Cos[n t])

FortranForm[%]

     -(c*o2*r**(-2 + k)*z**2*(k**2 + f*(k - n)*(k + n)*Cos(n*t)))/2.

bz4[r_,t_] = (o4)(z^4/24)trans2[r,t]

     1/24 o4 z^4 (-c f (-3+k) (-2+k) n^2 r^(-4+k) Cos[n t]+
     c (-3+k) (-2+k) k r^(-4+k) (1+f Cos[n t])+
     c (-3+k) (-2+k) (-1+k) k r^(-4+k) (1+f Cos[n t])+
     (-c f k n^2 r^(-2+k) Cos[n t]-c f (-1+k) k n^2 r^(-2+k) Cos[n t]+
     c f n^4 r^(-2+k) Cos[n t])/r^2+1/r (-c f (-2+k) n^2 r^(-3+k) Cos[n t]+
     c (-2+k) k r^(-3+k) (1+f Cos[n t])+
     c (-2+k) (-1+k) k r^(-3+k) (1+f Cos[n t])))

FullSimplify[%]

     1/24 c o4 r^(-4+k) z^4 ((-2+k)^2 k^2+f (-2+k-n) (k-n) (-2+k+n) (k+n) 
     Cos[n t])

FortranForm[%]

        (c*o4*r**(-4 + k)*z**4*
     -    ((-2 + k)**2*k**2 + 
     -      f*(-2 + k - n)*(k - n)*(-2 + k + n)*(k + n)*Cos(n*t)))/24.

(* Set the 1st and 3rd order terms of Br(z). *)
br1[r,t] = (o1)z D[b[r,t],{r,1}] 

     c k o1 r^(-1+k) z (1+f Cos[n t])

FullSimplify[%]

     c k o1 r^(-1+k) z (1+f Cos[n t])

FortranForm[%]

     c*k*o1*r**(-1 + k)*z*(1 + f*Cos(n*t))

br3[r,t] = -(o3)(z^3/6)D[trans[r,t],{r,1}]

     -(1/6) o3 z^3 (-c f (-2+k) n^2 r^(-3+k) Cos[n t]+
     c (-2+k) k r^(-3+k) (1+f Cos[n t])+
     c (-2+k) (-1+k) k r^(-3+k) (1+f Cos[n t]))

FullSimplify[%]

     -(1/6) c (-2+k) o3 r^(-3+k) z^3 (k^2+f (k-n) (k+n) Cos[n t])

FortranForm[%]

        -(c*(-2 + k)*o3*r**(-3 + k)*z**3*
     -     (k**2 + f*(k - n)*(k + n)*Cos(n*t)))/6.

(* Set the 1st and 3rd order terms of Btheta(z). *)
bt1[r,t] = (o1)z (1/r) D[b[r,t],{t,1}]

     -c f n o1 r^(-1+k) z Sin[n t]

FullSimplify[%]

     -c f n o1 r^(-1+k) z Sin[n t]

FortranForm[%]

     -(c*f*n*o1*r**(-1 + k)*z*Sin(n*t))

bt3[r,t] = -(o3)(z^3/6)(1/r)D[trans[r,t],{t,1}]

     -1/(6 r) o3 z^3 (-c f k n r^(-2+k) Sin[n t]-
     c f (-1+k) k n r^(-2+k) Sin[n t]+c f n^3 r^(-2+k) Sin[n t])

FullSimplify[%]

     1/6 c f (k-n) n (k+n) o3 r^(-3+k) z^3 Sin[n t]

FortranForm[%]

     (c*f*(k - n)*n*(k + n)*o3*r**(-3 + k)*z**3*Sin(n*t))/6.

\end{verbatim}

\section{Appendix:  {\tt FORTRAN} Example}

This program takes the output expressions of $B_z(z)$, $B_r(z)$, and 
$B_\theta(z)$ from Mathematica and prints out a grid of magnetic field values
recast as $B_x$, $B_y$, and $B_z$.  The output file generated by this routine
is fort.71.  With further formatting described in the initial comments of the
routine, the output file can be used by a G4Beamline input file to generate
a magnetic field for an anticyclotron simulation.
In this example, the field componenets are expanded to 4th order in $z$.

\begin{verbatim}

      PROGRAM FOURTH_ORDER_ARXIV_EXAMPLE

C          
C  Prints out grid of (Bx, By, Bz) in (x, y, z) for a file suitable for the
C  G4Beamline command 'fieldmap'.  
C
C  The output of this program is fort.71.  
C
C  Distances in the fort.71 file are in mm., and magnetic fields are 
C  in tesla. (Distances in this FORTRAN routine are in meters.) 
C
C  The following must be added to fort.71 for the file to work in G4Beamline:
C
C  grid X0=[smallest x] Y0=[smallest y] Z0=[smallest z] nX=[number of x points]
C nY=[number of y points] nZ=[number of z points] dX=[dist. between x points]
C dY=[dist. between y points] dZ=[dist. between z points]
C  data
C

      IMPLICIT NONE

      INTEGER I,J,L

      REAL*8 x,y,z,r,t
      REAL*8 bz0,bz2,bz4,br1,br3,bt1,bt3
      REAL*8 bx,by,bz,br,bt

      REAL*8 PI,E
      REAL*8 b,c,k,s,f,n

      REAL*8 o1,o2,o3,o4

      E = 2.718281828d+0
      PI = ACOS(-1.0d+0)

C  Bz(z=0) = (c*r^k)*(1+f*cos(n(t)))


C  Bz(z = 0) parameters
      b = 25.0d+0
      c = 1.77d+0
      k = 0.6d+0
      s = (PI/180.0d+0)*0.0d+0
      f = sqrt(2.0d+0)
      n = 6.0d+0

C  o1 through o4 set the order of the expansion in z of Bz(z=0)
C  4th order:  o4 = o3 = o2 = o1 = 1
C  3rd order:  o4 = 0, o3 = o2 = o1 = 1
C  2nd order:  o4 = o3 = 0, o2 = o1 = 1
C  1st order:  o4 = o3 = o2 = 0, o1 = 1
C  0th order:  o4 = o3 = o2 = o1 = 0

      o1 = 1.0d+0
      o2 = 1.0d+0
      o3 = 1.0d+0
      o4 = 1.0d+0


C  Loop over x, y, z
      DO 100 I = 1, 201
        DO 200 J = 1, 201
          DO 300 L = 1, 61

C  x, y, z, r in meters
C  t in radians
C  br, bt, bz in Tesla

            x = -1.01d+0 + 0.01d+0*I
            y = -1.01d+0 + 0.01d+0*J
            z = -0.31d+0 + 0.01d+0*L

            r = sqrt(x*x+y*y)

            t = atan2(y,x)

C  br, bt, bz formula determined through and copied and pasted from Mathematica

C  bz terms up to 4th order
            bz0 = c*r**k*(1 + f*Cos(n*t))

            bz2 = -(c*o2*r**(-2 + k)*z**2*(k**2 + 
     -             f*(k - n)*(k + n)*Cos(n*t)))/2.

            bz4 = (c*o4*r**(-4 + k)*z**4*
     -        ((-2 + k)**2*k**2 + 
     -        f*(-2 + k - n)*(k - n)*(-2 + k + n)*(k + n)*Cos(n*t)))/24.

C  br terms up to 4th order
            br1 = c*k*o1*r**(-1 + k)*z*(1 + f*Cos(n*t))

            br3 = -(c*(-2 + k)*o3*r**(-3 + k)*z**3*
     -       (k**2 + f*(k - n)*(k + n)*Cos(n*t)))/6.
            
C  bt terms up to 4th order
            bt1 = -(c*f*n*o1*r**(-1 + k)*z*Sin(n*t))

            bt3 = (c*f*(k - n)*n*(k + n)*o3*r**(-3 + k)*z**3*Sin(n*t))/6.

C  Add each order to get bz, br, bt
            bz = bz0 + bz2 + bz4

            br = br1 + br3

            bt = bt1 + bt3

C  Transform br, bt to bx, by
            bx=br*COS(t)-bt*SIN(t)

            by=br*SIN(t)+bt*COS(t)
            
            WRITE(71,174) 1000*x,1000*y,1000*z,bx,by,bz,0.0

174         FORMAT(F8.1,",",F8.1,",",F8.1,",",F9.3,",",F9.3,",",F9.3,
     1             ", 0.0, 0.0, ",F7.3)

 300      CONTINUE
 200    CONTINUE     
 100  CONTINUE   

      END

\end{verbatim}

Here are the first  10 lines in the fort.71 output file.

\begin{verbatim}

head fort.71
 -1000.0, -1000.0,  -300.0,   -3.318,    3.712,    2.162, 0.0, 0.0,   0.000
 -1000.0, -1000.0,  -290.0,   -3.160,    3.542,    2.163, 0.0, 0.0,   0.000
 -1000.0, -1000.0,  -280.0,   -3.007,    3.375,    2.164, 0.0, 0.0,   0.000
 -1000.0, -1000.0,  -270.0,   -2.859,    3.214,    2.165, 0.0, 0.0,   0.000
 -1000.0, -1000.0,  -260.0,   -2.716,    3.057,    2.166, 0.0, 0.0,   0.000
 -1000.0, -1000.0,  -250.0,   -2.576,    2.904,    2.167, 0.0, 0.0,   0.000
 -1000.0, -1000.0,  -240.0,   -2.441,    2.756,    2.168, 0.0, 0.0,   0.000
 -1000.0, -1000.0,  -230.0,   -2.310,    2.611,    2.169, 0.0, 0.0,   0.000
 -1000.0, -1000.0,  -220.0,   -2.182,    2.471,    2.170, 0.0, 0.0,   0.000
 -1000.0, -1000.0,  -210.0,   -2.058,    2.334,    2.171, 0.0, 0.0,   0.000

\end{verbatim}

With a grid spacing of 10 $mm$ ranging from $-1000~mm < (x,y) < 1000~mm$ and
$-300~mm < z < 300~mm$, the following lines need to be placed at the start of
\texttt{fort.71} to match the syntax required by the G4Beamline command \texttt{fieldmap}

\begin{verbatim}
param normB=1.0000 normE=0.0000
grid X0=-1000 Y0=-1000 Z0=-300 nX=201 nY=201 nZ=61 dX=10 dY=10 dZ=10
data
\end{verbatim}

\section{Appendix:  G4Beamline Example}

An example G4Beamline\,\cite{Roberts} input file of positive muons being decelerated in an 
anticycloton is shown.  The file
\texttt{test\_bfield\_6sector\_rk06\_0d\_2nd\_order\_cos\_example.dat}
as specified in the 
\texttt{fieldmap field\_grid file=test\_bfield\_6sector\_rk06\_0d\_2nd\_order\_cos\_example.dat} command
is needed to generate the outer sectored focusing field.  The inner magnetic
bottle is generated by the field from two current carrying coils.  Muons lose
energy by passing through $1.0^{\circ}$ lithium hydride wedges.  The uniform
$1.0^{\circ}$ wedge angle in this example file is simpler than the moderator 
configuration used for the anticyclotron.  The anticyclotron includes 
an inner cylinder of 0.1 $bar$ helium and six  
LiH wedges whose thickness decreases adiabatically from 4.2 $mm$ at 
$r = 550~mm$ to $0.008~mm$ at $r = 65~mm$.
The output is a root file, \texttt{AllTracks.root}, 
which shows all the kinematic information of each muon along its orbit,
and which can be opened by \texttt{Historoot}.

\begin{verbatim}
* example G4Beamline file for anticyclotron simulation, April 11, 2011
*
# use TJR recommended physics routines
physics QGSP_BERT doStochastics=1 disable=Decay

# reference particle with no stochastic processes
particlecolor reference=1,1,1
reference referenceMomentum=180 particle=mu+ \
    beamX=0.0 beamY=500.0 beamZ=0.0 meanXp=0.0 meanYp=0.0 rotation=X90,Z90

# simulate 3 muons with stochastic processes (since doStochastics=1)
beam gaussian particle=mu+ nEvents=3 \
beamX=0.0 beamY=500.0 beamZ=0.0 meanXp=0.0 meanYp=0.0 \
sigmaX=0 sigmaY=0 sigmaXp=0 sigmaYp=0 \
meanMomentum=180.00 sigmaP=0 meanT=0 sigmaT=0  rotation=X90,Z90
  
# keep only positive muons
trackcuts keep=mu+ maxTime=200000.0

# output file for Historoot showing kinematic information for orbits
trace nTrace=3 format=root filename=AllTracks.root oneNTuple=1 primaryOnly=1

# output file for Historoot showing kinematic information when particles are lost 
beamlossntuple loss_ntuple format=ascii

param maxStep=5.  SteppingVerbose=1

# lithium hydride, density (0.82 g/cm^3) from Wikipedia
material lih z=4 a=8 density=0.82 state=s

# Define LiH wedges, each with 1 degree angle
tubs gascylinder_1 innerRadius=2 outerRadius=600 length=200 \
  initialPhi=30-0.5 finalPhi=30+0.5 color=1,0,0 material=lih
tubs gascylinder_2 innerRadius=2 outerRadius=600 length=200 \
  initialPhi=90-0.5 finalPhi=90+0.5 color=1,0,0 material=lih
tubs gascylinder_3 innerRadius=2 outerRadius=600 length=200 \
  initialPhi=150-0.5 finalPhi=150+0.5 color=1,0,0 material=lih
tubs gascylinder_4 innerRadius=2 outerRadius=600 length=200 \
  initialPhi=210-0.5 finalPhi=210+0.5 color=1,0,0 material=lih
tubs gascylinder_5 innerRadius=2 outerRadius=600 length=200 \
  initialPhi=270-0.5 finalPhi=270+0.5 color=1,0,0 material=lih
tubs gascylinder_6 innerRadius=2 outerRadius=600 length=200 \
  initialPhi=330-0.5 finalPhi=330+0.5 color=1,0,0 material=lih

# Define coils of inner magnetic bottle
coil  coil1 \
	innerRadius=195 \
	outerRadius=205 \
	length=50 \
	material=Vacuum \
	filename=coil1.dat

coil  coil2 \
	innerRadius=195 \
	outerRadius=205 \
	length=50 \
	material=Vacuum \
	filename=coil2.dat

# Set currents for magnetic bottle coils which provide inner bottle magnetic field
solenoid  c1 \
        coilName=coil1 \
        current=(0.2)*10803.6 \
        color=0.3,1,0 \
        alternate=0

solenoid  c2 \
        coilName=coil1 \
        current=(0.2)*10803.6 \
        color=0.3,1,0 \
        alternate=0

# Grid of magnetic field points for outer sectored field
fieldmap field_grid file=test_bfield_no_poly_bottle_6sector_rk06_0d_2nd_order_beq25_cos.dat

# Place wedges, coils, and magnetic field map grid
place gascylinder_1 z=0
place gascylinder_2 z=0
place gascylinder_3 z=0
place gascylinder_4 z=0
place gascylinder_5 z=0
place gascylinder_6 z=0

place c1 x=0 y=0 z=200
place c2 x=0 y=0 z=-200
 
place field_grid x=0 y=0 z=0

\end{verbatim}

\begin{figure}[t!]
    \centering
    \includegraphics[width=100mm]{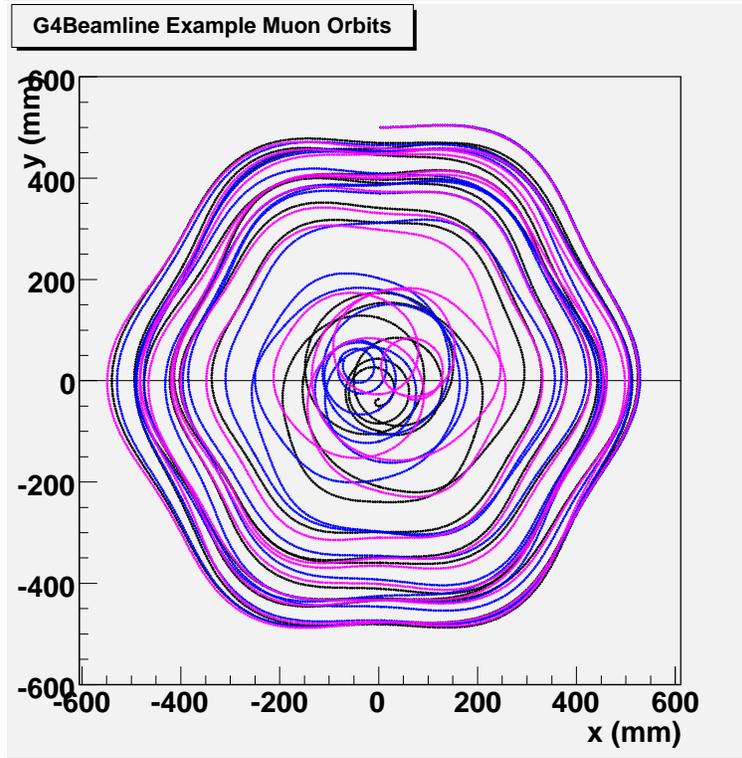}
    \vspace{-3mm}    
    \caption{Orbits of three color coded muons in an G4Beamline simulated 
             anticyclotron.  This output is generated by \texttt{Historoot}.  
             The orbits show the transition between the outer sectored field
             and the inner azimuthally symmetric magnetic bottle.}
    \label{g4bl_orbits}
\end{figure}

Fig.~\ref{g4bl_orbits} shows a \texttt{Historoot}
visualation of the orbits of three
muons in this G4Beamline example simulation.  The muons stop in about 125 $ns$.

\section{Appendix:  ICOOL Example}

Another program which can simulate an anticyclotron with 
lithium hydride wedges is ICOOL\,\cite{Fernow}.  The coordinate system of ICOOL is an 
$(x, y, z)$ coordinate system in which the $z$ coordinate is along the 
beamline.  This can be converted to an $(r, \theta, z)$ system by 
incorporating a bent solenoid so that $x$ becomes $-r$, $y$ becomes $z$, and 
the reference path coordinate, $z$, becomes $\theta$.  ICOOL does not have
the visualization capabilites of G4Beamline so software to check element placement, orbits,
and dynamics must be added.  ICOOL can do its own magnetic field expansion\,\cite{bent-sol} to check an orbit at a particular radius.
This can provide a cross check of  G4Beamline.

ICOOL can be run on UNIX, PC, and Macintosh platforms.  
The command file needs to be called
\texttt{for001.dat}.  
In this particular application of ICOOL, two more input files
are used:  \texttt{for003.dat} which defines each particle of the beam and
\texttt{for056.dat} which defines the magnetic field.  The output file
in the example simulation is \texttt{for009.dat} which lists the
kinematic information of each particle.  Postprocessing of this file is
needed to visualize the orbits and to evaluate the dynamics.

Information about ICOOL including a User's Guide and Reference Manual can be 
found at 
\begin{verbatim}
  https://pubweb.bnl.gov/~fernow/icool/v326/
\end{verbatim}
The next
sections show example input files which generate an anticyclotron 
simulation when the executable \texttt{icool} from
\begin{verbatim}
  https://pubweb.bnl.gov/~fernow/icool/v326/icool.exe
\end{verbatim}
is run.

\subsection{input for001.dat command file}
This command files uses a bent solenoid magnetic field to simulate a
cyclotron field.  The \texttt{BSOL} command sets the initial muon momentum
to 0.18 $GeV/c$, the inverse radius of the 
bent solenoid to 2.0 $m^{-1}$, and reads the amplitude, period, and initial 
offset
of the sinusoidal vertical magnetic field ($B_y(z)$ where $y$ is vertical and
$z$ is along the reference path) from the file \texttt{fort.56}.  In this
ICOOL example, muons lose energy by passing through $1.0^\circ$ 
lithium hydride wedges.  The magnetic field in this example does not vary
with radius is in the G4Beamline example of Appendix C.  
The geometry of the anticyclotron is specified by \texttt{sregions}.
One \texttt{sregion} of the lithium hydride wedge is followed by 10 
\texttt{sregions} of vacuum, and the sequence of 11 \texttt{sregions} is
repeated 600 times.  The total length of the 11 \texttt{sregions} is 
$0.523598775~m$ which corresponds to 1/6 of the circumference of a $0.5~m$
circle.  This arrangement is set for a 6-sector anticyclotron with a 
bent solenoid radius of $0.5~m$. 
\begin{verbatim}

test ring 180 MeV (H5b)

$cont npart=200 nsections=1 varstep=.false. nprnt=-3 prlevel=1 
ntuple=.false. rtuple=.false. rtuplen=1 output1=.true. phasemodel=3
fsav=.false. fsavset=.true. izfile=140 bgen=.false.  $

 $bmt $
1 2 1. 1              !ntyp typ(1e 2mu 3pi 4K 5p) frac dist(1=gauss 2=ring)
-0.010  0.019   0.  0.   0.   0.203  !means: x y z px py pz
0.0425  0.0425 .18 .030 .030  0.024    !sigmas:     "
1                                    !correlations
2  .19  .40  0                       !palmer  GeV/A^2 beta

 $ints ldecay=.false.  declev=1  ldedx=.true. lstrag=.true. lscatter=.true.
  delev=2 straglev=4 scatlev=4  $

 $nhs  $
 $nsc  $
 $nzh  $
 $nrh nrhist=0 $
 $nem  $
 $ncv  $
 
SECTION

REFP
2  .18  0. 0. 3             !typ refp t0 grad0 mode (3=const p 4=with acc)

DENS
LIH 1
BEGS

!=========================================================================

CELL      !-------------------- regular cell 
1    
.FALSE.

BSOL                                    ! multipole field input
4. 56 .180 5 1 1 1 1 1 1 1 1 1 1 2.0    ! mode, file, momentum, order, 1/r switch, 10 scale-factors, 1/r

!----------------------------------------------------------------------

REPEAT
1200

OUTPUT
SREGION                       ! define a region              
.023598775 1 0.0001                   ! length, 1 radial subregion, step
1 0. 2                        ! radial extent
NONE                          ! no associated field
0. 0. 0. 0. 0. 0. 0. 0. 0. 0. 0. 0. 0. 0. 0.
LIH VAC                       ! LiH material surrounded by vacuum
WEDGE
1 0.5 0.01 180 1.0 1.5 0. 0. 0. 0.      !angle RVERT ZVERT Azymuth Apx Apy

OUTPUT
SREGION                       ! define a region              
.05 1 0.001            ! length, 1 radial subregion, step
1 0. 2                        ! radial extent
NONE                          ! no associated field
0. 0. 0. 0. 0. 0. 0. 0. 0. 0. 0. 0. 0. 0. 0.
VAC                           ! vacuum
CBLOCK                        ! cylindrical block geometry
0. 0. 0. 0. 0. 0. 0. 0. 0. 0.

OUTPUT
SREGION                       ! define a region              
.05 1 0.001            ! length, 1 radial subregion, step
1 0. 2                        ! radial extent
NONE                          ! no associated field
0. 0. 0. 0. 0. 0. 0. 0. 0. 0. 0. 0. 0. 0. 0.
VAC                           ! vacuum
CBLOCK                        ! cylindrical block geometry
0. 0. 0. 0. 0. 0. 0. 0. 0. 0.

OUTPUT
SREGION                       ! define a region              
.05 1 0.001            ! length, 1 radial subregion, step
1 0. 2                        ! radial extent
NONE                          ! no associated field
0. 0. 0. 0. 0. 0. 0. 0. 0. 0. 0. 0. 0. 0. 0.
VAC                           ! vacuum
CBLOCK                        ! cylindrical block geometry
0. 0. 0. 0. 0. 0. 0. 0. 0. 0.

OUTPUT
SREGION                       ! define a region              
.05 1 0.001            ! length, 1 radial subregion, step
1 0. 2                        ! radial extent
NONE                          ! no associated field
0. 0. 0. 0. 0. 0. 0. 0. 0. 0. 0. 0. 0. 0. 0.
VAC                           ! vacuum
CBLOCK                        ! cylindrical block geometry
0. 0. 0. 0. 0. 0. 0. 0. 0. 0.

OUTPUT
SREGION                       ! define a region              
.05 1 0.001            ! length, 1 radial subregion, step
1 0. 2                        ! radial extent
NONE                          ! no associated field
0. 0. 0. 0. 0. 0. 0. 0. 0. 0. 0. 0. 0. 0. 0.
VAC                           ! vacuum
CBLOCK                        ! cylindrical block geometry
0. 0. 0. 0. 0. 0. 0. 0. 0. 0.

OUTPUT
SREGION                       ! define a region              
.05 1 0.001            ! length, 1 radial subregion, step
1 0. 2                        ! radial extent
NONE                          ! no associated field
0. 0. 0. 0. 0. 0. 0. 0. 0. 0. 0. 0. 0. 0. 0.
VAC                           ! vacuum
CBLOCK                        ! cylindrical block geometry
0. 0. 0. 0. 0. 0. 0. 0. 0. 0.

OUTPUT
SREGION                       ! define a region              
.05 1 0.001            ! length, 1 radial subregion, step
1 0. 2                        ! radial extent
NONE                          ! no associated field
0. 0. 0. 0. 0. 0. 0. 0. 0. 0. 0. 0. 0. 0. 0.
VAC                           ! vacuum
CBLOCK                        ! cylindrical block geometry
0. 0. 0. 0. 0. 0. 0. 0. 0. 0.

OUTPUT
SREGION                       ! define a region              
.05 1 0.001            ! length, 1 radial subregion, step
1 0. 2                        ! radial extent
NONE                          ! no associated field
0. 0. 0. 0. 0. 0. 0. 0. 0. 0. 0. 0. 0. 0. 0.
VAC                           ! vacuum
CBLOCK                        ! cylindrical block geometry
0. 0. 0. 0. 0. 0. 0. 0. 0. 0.

OUTPUT
SREGION                       ! define a region              
.05 1 0.001            ! length, 1 radial subregion, step
1 0. 2                        ! radial extent
NONE                          ! no associated field
0. 0. 0. 0. 0. 0. 0. 0. 0. 0. 0. 0. 0. 0. 0.
VAC                           ! vacuum
CBLOCK                        ! cylindrical block geometry
0. 0. 0. 0. 0. 0. 0. 0. 0. 0.

OUTPUT
SREGION                       ! define a region              
.05 1 0.001            ! length, 1 radial subregion, step
1 0. 2                        ! radial extent
NONE                          ! no associated field
0. 0. 0. 0. 0. 0. 0. 0. 0. 0. 0. 0. 0. 0. 0.
VAC                           ! vacuum
CBLOCK                        ! cylindrical block geometry
0. 0. 0. 0. 0. 0. 0. 0. 0. 0.


ENDREPEAT

ENDCELL
ENDSECTION

\end{verbatim}

\subsection{input for003.dat beam file}
This input file defines the initial positions and momenta of three input
muons.  The initial positions are set to zero with respect to the 0.5 m
reference radius of the bent solenoid field.  The initial momentum of 
the three muons is 0.18 $GeV/c$ in the $z$ direction.
\begin{verbatim}

title
0  0  0  0  0  0  0  0 
    1 0 2 0   0.0000E+00 1.0
0.00E-00  0.0000E+00  0.0000E+00 0.0000E-03  0.0000E+00  1.8000E-01 0 0 0 0 0 0
    2 0 2 0   0.0000E+00 1.0
0.00E-00  0.0000E+00  0.0000E+00 0.0000E-03  0.0000E+00  1.8000E-01 0 0 0 0 0 0
    3 0 2 0   0.0000E+00 1.0
0.00E-00  0.0000E+00  0.0000E+00 0.0000E-03  0.0000E+00  1.8000E-01 0 0 0 0 0 0


\end{verbatim}

\subsection{input for056.dat field file}
This file sets the parameters of the bent solenoid field.  This field
is sinusoidal with respect to the $z$ coordinate (In ICOOL, the
vertical component is $y$ and $z$ is the coordinate along the 
reference path.).  The period is about
0.5236 $m$ corresponding to 1/6 the circumference of a circle with reference
radius of 0.5 $m$.  The field is independent with respect to
radius and depends on the
$z$ coordinate as $B_y(z) = 1.168 - 1.651\cos(2\pi z/0.5236)$ where the
field is in Tesla and the distance is in meters.
\begin{verbatim}

test 6 sector anticyclotron
 0.523598775  1  1  1  1  1  1  1  1  1  1  1  1  1
1
n     s           d     a          q
 0  0  0 1.167764501  0  0  0  0  0  0  0  0  0  0  0  0  0  0  0  0  0  0  0  
 0  0  0  0
 1  0  0 -1.651468395  0  0  0  0  0  0  0  0  0  0  0  0  0  0  0  0  0  0  
 0  0  0  0  0

\end{verbatim}

\subsection{output for009.dat file}

The following shows the kinematic information of the reference muon and the
three muons through one wedge sregion and one vacuum sregion.  Each data line 
wraps around to form three actual lines. 

\begin{verbatim}
#test ring 180 MeV (H5b)                                                        
#  units = [s] [m]  [GeV/c] [T] [V/m] 
#   evt par typ  flg reg  time      x        y       z 
  Px         Py      Pz       Bx       By      Bz     wt Ex  Ey Ez arc 
  polX polY polZ
      0  0   2    0   1  0.0E+00 0.0E+00  0.0E+00 0.0E+00 
 0.0E+00  0.0E+00 1.8E-01  0.0E+00  0.0E+00  0.0E+00  1   0  0  0 0.0E+00 
   0    0    0
      1  0   2    0   1  0.0E+00 0.0E+00  0.0E+00 0.0E+00 
 0.0E+00  0.0E+00 1.8E-01  0.0E+00  0.0E+00  0.0E+00  1   0  0  0 0.0E+00 
   0    0    0
      2  0   2    0   1  0.0E+00 0.0E+00  0.0E+00 0.0E+00 
 0.0E+00  0.0E+00 1.8E-01  0.0E+00  0.0E+00  0.0E+00  1   0  0  0 0.0E+00 
   0    0    0
      3  0   2    0   1  0.0E+00 0.0E+00  0.0E+00 0.0E+00 
 0.0E+00  0.0E+00 1.8E-01  0.0E+00  0.0E+00  0.0E+00  1   0  0  0 0.0E+00 
   0    0    0
      0  0   2    0   5  9.1E-11 0.0E+00  0.0E+00 2.4E-02 
 0.0E+00  0.0E+00 1.8E-01  0.0E+00 -4.2E-01  0.0E+00  1   0  0  0 2.4E-02 
   0    0    0
      1  0   2    0   5  9.2E-11 7.8E-04  1.2E-04 2.4E-02 
 1.1E-02  1.9E-03 1.7E-01  2.4E-10 -4.2E-01  6.4E-04  1   0  0  0 2.4E-02 
   0    0    0
      2  0   2    0   5  9.2E-11 1.1E-03 -7.4E-05 2.4E-02 
 1.5E-02 -1.7E-03 1.7E-01 -6.2E-11 -4.2E-01 -4.1E-04  1   0  0  0 2.4E-02 
   0    0    0
      3  0   2    0   5  9.2E-11 7.8E-04 -1.6E-04 2.4E-02 
 1.2E-02 -1.5E-03 1.8E-01 -6.2E-10 -4.2E-01 -8.8E-04  1   0  0  0 2.4E-02 
   0    0    0
\end{verbatim}

\begin{figure}[h!]
    \centering
    \includegraphics[width=100mm]{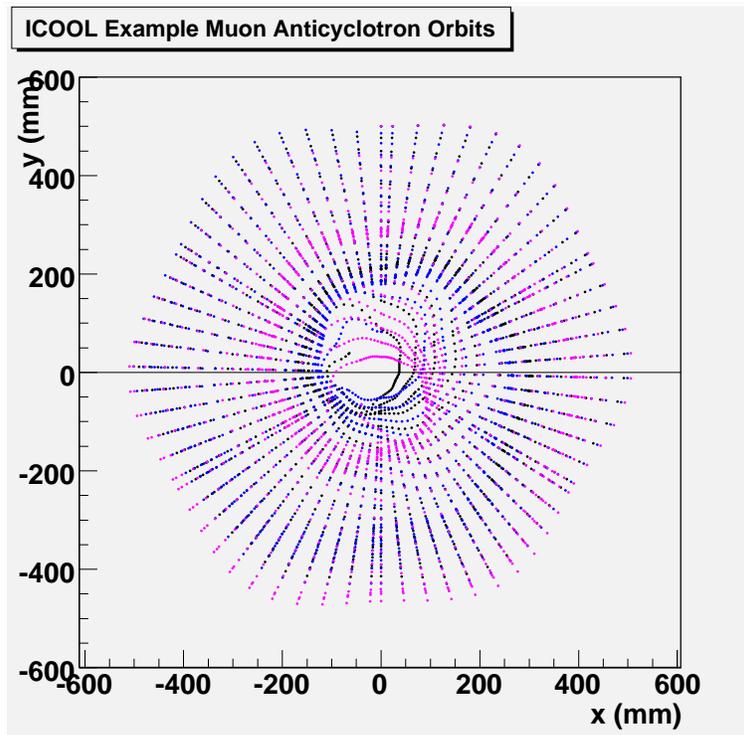}
    \vspace{-3mm}    
    \caption{Orbits of three color coded muons in an ICOOL simulated 
             anticyclotron.  The output is generated by \texttt{Historoot}.
             Each
             point in the orbit is at the location of the end of an sregion.
             Unlike the G4Beamline simulation, the magnetic field in this 
	     ICOOL simulation is independent of radius and
             does not include an
             inner magnetic bottle.}
    \label{icool}
\end{figure}

The orbit in the LAB $(x,y)$ plane can be viewed with \texttt{Historoot} which
can be obtained by downloading \texttt{G4Beamline} from 
\texttt{www.muonsinc.com}.  With \texttt{for009.dat} as the input file to 
\texttt{Historoot}, the orbit in the LAB $(x,y)$ frame is obtained by plotting
$(x+0.5~m)\cos(z/0.5~m)$ vs. $(x+0.5~m)\sin(z/0.5~m)$.  
The radius is the LAB frame
of the anticyclotron is $(x+0.5~m)$ where $x$ is the the horizontal coordinate
with respect to the ICOOL reference path which has a
bent solenoid radius
of curvature $0.5~m$.  
Fig.~\ref{icool} shows the orbits of three muons in this
example ICOOL anticyclotron simulation which are stopped in about 200 $ns$.

\end{document}